\documentclass[conference]{IEEEtran}
\IEEEoverridecommandlockouts

\usepackage{amsmath,amssymb,amsfonts}
\usepackage{graphicx}
\usepackage{textcomp}
\usepackage{xcolor}
\usepackage{threeparttable}
\usepackage{multirow}
\usepackage{threeparttable}
\usepackage{multirow}
\usepackage{makecell} 
\usepackage{multicol}
\usepackage{caption}
\usepackage{subcaption}
\usepackage{adjustbox}
\usepackage{pifont}
\usepackage{algorithm}
\usepackage[noend]{algpseudocode}
\usepackage{url}
\usepackage{amsthm}
\usepackage{xspace}

\newcommand{\specialcell}[2][c]{
\begin{tabular} [#1]{@{}c@{}}#2\end{tabular}}

\newcommand{\cmark}{\ding{51}}%
\newcommand{\xmark}{\ding{55}}%
\newtheorem{theorem}{Theorem}

\newtheorem{corollary}{Corollary}
\newcommand{\Ra}{\ensuremath \stackrel{\$}{\leftarrow}{\xspace}}
\newcommand{\as}{\ensuremath {\leftarrow}{\xspace}}

\newcommand{\prf}{\ensuremath {\texttt{PRF}}{\xspace}}
\newcommand{\oo}{\ensuremath {\texttt{OO}}{\xspace}}
\newcommand{\enc}{\ensuremath {\texttt{ENC}}{\xspace}}
\newcommand{\encoo}{\ensuremath {\texttt{ENC-OO}}{\xspace}}

\newcommand{\encmac}{\ensuremath {\texttt{EncMac}}{\xspace}}
\newcommand{\encstddec}{\ensuremath {\texttt{ENC-Std.Dec}}{\xspace}}

\newcommand{\A}{\ensuremath {\mathcal{A}}{\xspace}}

\newcommand{\encstd}{\ensuremath {\texttt{ENC-Std}}{\xspace}}
\newcommand{\encstdenc}{\ensuremath {\texttt{ENC-Std.Enc}}{\xspace}}

\newcommand{\macoo}{\ensuremath {\texttt{MAC-OO}}{\xspace}}
\newcommand{\macoomac}{\ensuremath {\texttt{MAC-OO.Mac}}{\xspace}}
\newcommand{\macstd}{\ensuremath {\texttt{MAC-Std}}{\xspace}}
\newcommand{\macstdmac}{\ensuremath {\texttt{MAC-Std.Mac}}{\xspace}}
\newcommand{\macaver}{\ensuremath {\texttt{MAC.AVer}}{\xspace}}

\newcommand{\bmacoo}{\ensuremath {b_\texttt{MAC-OO}}{\xspace}}
\newcommand{\bencoo}{\ensuremath {b_\texttt{ENC-OO}}{\xspace}}
\newcommand{\bagg}{\ensuremath {b_\texttt{AGG}}{\xspace}}
\newcommand{\bbver}{\ensuremath {b_\texttt{BVer}}{\xspace}}

\newcommand{\sk}{\ensuremath { \mathit{sk} }{\xspace}}

\newcommand{\SK}{\ensuremath { \boldsymbol{\mathit{sk}} }{\xspace}}

\newcommand{\mac}{\texttt {MAC}{\xspace}}

\newcommand{\graphene}{\ensuremath {\texttt{Graphene}}{\xspace}}
\newcommand{\grapheneoo}{\ensuremath {\texttt{Graphene.OO}}{\xspace}}
\newcommand{\graphenekg}{\ensuremath {\texttt{Graphene}\texttt{.Kg}}{\xspace}}
\newcommand{\grapheneupd}{\ensuremath {\texttt{Graphene}\texttt{.Upd}}{\xspace}}
\newcommand{\grapheneagg}{\ensuremath {\texttt{Graphene}\texttt{.Agg}}{\xspace}}
\newcommand{\grapheneencmac}{\ensuremath {\texttt{Graphene}\texttt{.EncMac}}{\xspace}}
\newcommand{\grapheneverdec}{\ensuremath {\texttt{Graphene}\texttt{.VerDec}}{\xspace}}

\newcommand{\graphenepoly}{\ensuremath {\texttt{Graphene-Poly}}{\xspace}}

\newcommand{\grapheneae}{\ensuremath {\texttt{Graphene-AE}}{\xspace}}

\newcommand{\algrule}{\par\vskip.4\baselineskip\hrule\vskip.4\baselineskip\par}

\usepackage[
top    = 0.76in,
bottom = 1.04in,
left   = 0.625in,
right  = 0.625in
]{geometry}
\begin{document}
\title{Lightweight and Breach-Resilient Authenticated Encryption Framework for Internet of Things}
\author{
    Saif Eddine Nouma, 
    Attila A. Yavuz \\[1ex]
    University of South Florida, Tampa, FL, USA \\
    Email: \{saifeddinenouma, attilaayavuz\}@usf.edu
}
\maketitle

\begin{abstract}
The Internet of Things (IoT) relies heavily on resource-limited devices to communicate critical (e.g., military data) information under low-energy adversarial environments and low-latency wireless channels. Authenticated Encryption (AE) guarantees confidentiality, authenticity, and integrity, making it a vital security service for IoT. However, current deployed (lightweight) AE standards lack essential features like key compromise resiliency and compact authentication tags, as well as performance enhancements such as offline-online cryptography. To address these gaps, we propose \graphene, the first (to our knowledge) symmetric Forward-secure and Aggregate Authenticated Encryption (FAAE) framework designed for the performance and security demands of low-end IoT infrastructures. \graphene~innovates by synergizing key evolution strategies and offline-online cryptographic processing with Universal Message Authentication Codes (UMACs) to guarantee breach-resiliency, near-optimal online latency, and compactness. We demonstrate \graphene's efficiency through two distinct instantiations, each balancing unique performance trade-offs with extensibility for diverse MACs. Our experimental evaluation on commodity hardware and 32-bit ARM Cortex-M4 microcontroller shows \graphene's significant performance gains over existing alternatives. \graphene~is also backward compatible with standard-compliant cryptographic implementations. We release our implementation as open source for public testing and adaptation.
\end{abstract}


\section{INTRODUCTION} \label{sec:introduction}
The rapid expansion of wireless IoT systems across critical domains like smart healthcare, battlefield surveillance, and industrial automation has made secure communication a critical necessity \cite{wagner2022bp, SUHaSAFSS11, hiller2018secure, dubrova2018lightweight}. Resource-constrained IoT devices, such as medical pacemakers, aerial drones, and surveillance cameras, continuously generate sensitive telemetry and offload it to an edge or cloud server for remote diagnosis and real-time alerts \cite{yaqoob2019security}. However, ensuring secure and efficient wireless communication remains challenging due to the inherent resource constraints of embedded devices, including limited computational power, bandwidth, and battery life.

Authenticated Encryption (AE) \cite{degabriele2024sok} combines symmetric encryption with Message Authentication Codes (MACs) to provide confidentiality, integrity, and authenticity, making it a compelling solution for wireless IoT infrastructures. While existing AE standards (e.g., AES-GCM \cite{mcgrew2004security}) achieve high efficiency through hardware acceleration across various platforms \cite{hofemeier2012introduction}, they fall short of fully utilizing advanced cryptographic techniques to meet these rigorous security and performance demands of constrained IoT networks:
\textbf{\em (1) Offline-Online (OO) cryptography} to enable low-latency and energy-efficient processing \cite{wagner2022bp, hiller2018secure}.
\textbf{\em (2) Integration of Universal MACs within AEs}, leveraging partial homomorphism \cite{etzel1999square, le2022efficient} for efficient batch verification of authentication tags.
\textbf{\em (3) Forward security} to mitigate the impact of key compromise attacks \cite{SUHaSAFSS11, FssAggMAC_DiMa}.
\textbf{\em (4) Tag aggregation} to reduce transmission and storage overhead, enhancing overall system efficiency \cite{wagner2024and, eikemeier2010history}. In the following sections, we review related work on AE and MAC schemes, with a particular emphasis on these essential properties.

\subsection{Related Work and Limitations}
ASCON~\cite{li2023automatic}, the NIST lightweight AE standardization winner, lacks properties (1)-(4) due to its integrated AE structure. Specifically, its encryption precludes ciphertext-independent preprocessing while tag verification requires full payload decryption, making it incompatible with desirable efficient batch verification. It is also vulnerable to preimage attacks \cite{li2023automatic}, which significantly constrain its use in critical domains.

AES-GCM \cite{hiller2018secure} is a widely used AE standard in protocols like DTLS 1.3 and IEEE 802.15.4  for constrained sensor and low-rate wireless personal area networks. ChaCha20-Poly1305 \cite{degabriele2024sok} is also used in TLS 1.3 for IoT chips (e.g., ARM Cortex-M) when hardware-accelerated AES is not available. Both share the same design principle: a counter mode of operation (CTR) for encryption, while authentication is based on a Wegman-Carter MAC construction \cite{degabriele2024sok} from polynomial hash functions. 

Few encryption works explore precomputation (1) of the counter mode, leveraging idle periods of low-end devices. For instance, \cite{hiller2018secure} analyzes latency in industrial applications and proposes precomputed block cipher templates.

Wegman-Carter MAC constructions \cite{degabriele2024sok} are more efficient than traditional Hash-based MACs (HMACs) \cite{SUHaSAFSS11}. They offer precomputation (1) due to nonce-based structure. To our knowledge, none have explored AE integration with desirable properties (1-4). GHASH and Poly1305, as authenticators of AES-GCM and ChaCha20-Poly1305 provide (1) and amenable to key update (3) and tag aggregation (4). \cite{degabriele2024sok} proposes variants of Poly1305 with security and performance trade-offs. \cite{wagner2022bp} supports (1) but is efficient only for small inputs ($\approx 1$-$4$ bytes), while including forward security is costly due to large pre-stored key tables. Some MAC variants based on universal hash functions \cite{etzel1999square, le2022efficient} provide efficient batch verification (2). 

Forward-secure and aggregate MACs \cite{FssAggMAC_DiMa, SUHaSAFSS11} achieve compact aggregate tags and key compromise resiliency but omits precomputation (1) along with an integrated precomputable encryption. Progressive MACs \cite{wagner2022take} offers short authentication tags and provides resiliency against synchronization attacks, but with tricky internal state management and lacks (2-4).

Overall, there is a gap in the state of the art in achieving lightweight, breach-resilient, precomputable, and compact authenticated encryption for resource-constrained IoT systems.  

\subsection{Our Contributions}
We propose \graphene, to our knowledge, the first comprehensive breach-resilient and compact AE framework to enhance resiliency and online performance via forward-security, pre-computation, partial homomorphism with batch processing, and various aggregation modes. We outline the desirable properties of \graphene~as follows: \vspace{1pt}

\noindent \textbf{\em $\bullet$ A New Forward-secure and Aggregate AE~Framework (FAAE)}: \sloppy \graphene~harnesses key evolution with sequential tag aggregation considering special MACs and combination modes. \graphene~uncovers synergies among forward-secure encryption and UMACs overlooked by previous approaches while being backward compatible with current standard implementations. Therefore, it achieves high efficiency and security while opening a path for alternative constructions through generalizable integrations and ease of implementation. \vspace{1pt}

\noindent \textbf{\em $\bullet$ High Computation Efficiency via OO Cryptography}:  
OO cryptography shifts input-independent cryptographic computations to the pre-processing stage during idle periods, reducing online latency. Although studied for digital signatures, OO methods have not been explored for {FAAE}. \graphene~explores the synergies of various UMACs and AE modes in OO settings and achieves a significant online performance gain with only a modest increase in memory consumption. For example, our \graphenepoly~instantiation with OO is {\em $28\times$ faster than} FAAE in a batch of 1024 16-byte telemetry on a 32-bit Cortex-M4 embedded processor, with only 32KB extra storage (12\% of available SRAM). In wearable medical settings, this results in significant savings in battery life during online operations, as the next batch of OO keys can be supplied when the device is recharged while telemetry is uploaded. In aerial drone applications, latency reduction can improve flight safety and real-time telemetry transmission (where key storage is a lesser concern). Finally, beyond benefiting resource-limited devices, fast online operations are advantageous for receivers (e.g., edge clouds) that may need to verify and decrypt millions of telemetry packets simultaneously from various senders. ~\vspace{1pt}  

\noindent \textbf{\em $\bullet$ Adaptable Performance Choices and Tunable Security with \graphene~Instantiations}: \graphene~instantiations integrate well-established standards and MACs, offering enhanced performance and security features: {\em (i)} \grapheneae~provides a standard security level (128-bit) and compliance, operating $3\times$ faster than basic FAAE with an additional 16KB storage on 16-byte inputs with a single-key initialization. {\em (ii)} \graphenepoly~delivers high efficiency with medium 103-bit security, running $8.1\times$ faster than \grapheneae~with the same 16KB storage requirement. 
\graphene~supports aggregation in various modes, including cryptographic hash functions and XORing, each providing distinct properties.
\vspace{1pt}

\noindent \textbf{\em $\bullet$ Full-Fledged Implementation and Comprehensive Benchmarks}: \sloppy We implemented \graphene~on commodity hardware (x86/64) and a low-end microcontroller (32-bit ARM Cortex-M4). Our performance results highlight the efficiency of each instantiation, guiding its application context accordingly. To encourage reproducibility, we release our implementation at: 
\vspace{1pt}

\noindent \fbox{\url{https://github.com/saifnouma/Graphene}}

\section{Preliminaries and Models}
\label{sec:preliminaries}

\noindent \textbf{Notation.}
$||$ and $|x|$ denote concatenation and bit length of variable $x$, respectively.
$x \Ra \mathcal{S}$ denotes $x$ is randomly selected from the finite set $\mathcal{S}$ using a uniform distribution.
$\{0,1\}^*$ denotes a set of binary strings of any finite length. $\mathbb{Z}_q$ denotes the finite field of prime order $q$.
$\boldsymbol{x}$ denotes a set of items $\{x_1,\ldots,x_n\}$, where $n=|\boldsymbol{x}|$.
 $\oplus$ denotes XOR operation. $Add_q$ denotes modular addition over modulus $q$.  
\prf~is a keyed pseudo-random function. It is secure if its output ($\prf_k(m)$) is indistinguishable from a random string.
$H$ is a collision-resistant one-way cryptographic hash function.
 \vspace{2pt}

\noindent \textbf{Universal MACs.} \cite{etzel1999square} consist of a Carter-Wegman construction with a universal hash family for efficient data integrity. Formally, $\mac_{k}(m, n) = F_r(m) + \prf_{s}(n)$ where $k\as(r,s)$ is the private key, $F$ is a universal hash family and $\prf_{s}(n)$ is precomputable. $m$ and $n$ are the message and a counter, respectively. 
The security of universal MACs depends on the security of \prf~and $F$. $F$ is secure if it is $\epsilon$-almost universal hash function, i.e., $Pr[F_{r}(m)-F_r(m')=\delta] = \epsilon$ where $(m,m')$ are distinct messages and $\delta$ in the range of $F$. 
\vspace{2pt}

\begin{figure}[ht!]
	\centering
	\includegraphics[scale=0.46]{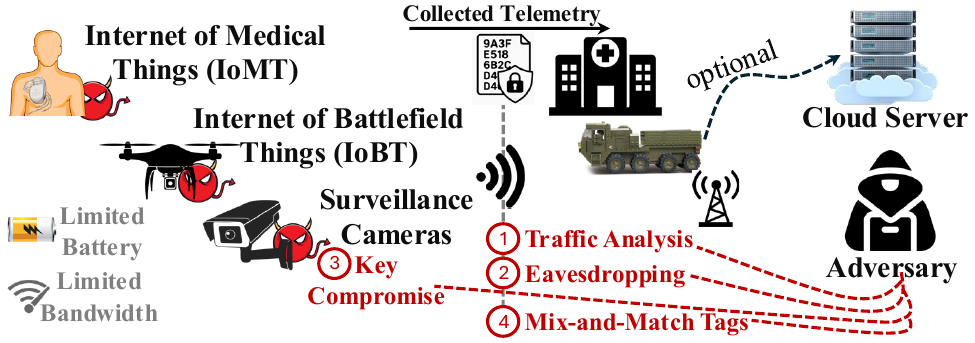}
	\caption{Our system and threat models}
	\label{fig:systemmodel}
	\vspace{-4pt}
\end{figure}

\noindent \textbf{System Model.}
 It consists of two main entities, as depicted in Fig. \ref{fig:systemmodel}: 
{\em (1) Resource-constrained IoT devices (e.g., medical implants)}: operate in adversarial environments under constrained resources of storage, memory, and computation. They continuously generate sensitive data and then periodically upload the telemetry to a nearby edge server. This storage-and-forward setting has vast applications in IoT, such as digital twins and wearable medical devices~\cite{yaqoob2019security, nouma2024trustworthy}.  
{\em (2) Resourceful verifier}: is a storage and computation resourceful entity, being a patient's phone or a clinical gateway in IoMT, or a nearby access point in IoBT. It receives encrypted and authenticated data from low-end IoT devices, verifies and decrypts them, and optionally offloads to a cloud server for diagnosis and actuation.  
The two main entities pre-share a private key. \vspace{2pt}

\noindent \textbf{Threat Model.} We assume a resourceful probabilistic polynomial-time (PPT) adversary $\mathcal{A}$ capable of launching: 
{\em (1) Passive attacks} monitor and analyze encrypted and authenticated traffic.
{\em (2) Active attacks} manipulate encrypted packets during transmission. 
{\em (3) Key compromise attacks} extract the long-term symmetric private keys during a system breach (e.g. malware) or a side-channel attack.
{\em (4) Mix-and-match attacks} target systems that mix multiple aggregate tags across multiple users to forge an aggregate MAC~scheme \cite{eikemeier2010history}.
The adversary aims to decrypt previously intercepted traffic using compromised private keys and produce a forgery on the aggregate and forward-secure authentication tag with the stored telemetry. \vspace{2pt}

 \noindent \textbf{Security Model.} Our security model is based on the security of its underlying (authenticated) encryption and MAC components. We consider forward security, which implies periodic key updates, an essential feature in adversarial IoT environments against key compromise attacks.  For confidentiality, \graphene~adopts forward-secure indistinguishability under chosen plaintext attacks (F-IND-CPA) \cite{mcgrew2004security}. For MACs, we follow the security of previous FAAEs  \cite{SUHaSAFSS11, FssAggMAC_DiMa}, with forward-secure existential unforgeability against chosen message attacks (F-EU-CMA) \cite{nouma2023post}. F-EU-CMA extends to FA-EU-CMA to cover security against mix-and-match attacks against aggregate MACs \cite{eikemeier2010history}. These notions collectively ensure that even an adaptive PPT adversary $\mathcal{A}$ can neither decrypt previously and currently encrypted telemetry traffic nor forge valid aggregate and forward-secure tags without detection.  \A~$(t,\epsilon, q_s)$-break denotes \A~can break a cryptographic scheme in time $t$ with probability $\epsilon$ and at most $q_s$ queries.



\section{PROPOSED SCHEMES} \label{sec:schemes}

A standard FAAE scheme typically applies conventional encryption with an HMAC. Its aggregation and forward security mechanisms are built on only a standard hash function. However, this construction falls short of achieving the high efficiency and desirable properties outlined in Section \ref{sec:introduction}, such as precomputation OO capabilities for minimal online latency, partial homomorphic features, and flexible aggregation modes, all essential for our system model. While prior MACs have explored precomputation, these efforts were isolated, neglecting key evolution and encryption. In this work, we introduce \graphene, the first FAAE framework designed to deliver near-optimal online computational efficiency and compactness, tailored for resource-constrained IoT environments.

\begin{figure}[ht!]
	\centering
	\noindent 
        \fbox{
        \parbox{0.97\columnwidth} {
			\scriptsize
			
			\begin{algorithmic}[1]
				\Statex \underline{$\SK_1 \as \graphenekg(1^\kappa,n,\bencoo,\bmacoo,\bagg,\bbver)$}: $n$~denotes batch size.  (\bencoo, \bmacoo) indicate if \enc~and \mac~admits \oo, respectively. (\bagg, \bbver) denote aggregation mode and batch verification capability, respectively. 
				
				\State Given security parameter $\kappa$, generate the initial (root) key $\SK_1 \as (\sk_1,\sk_1')$, where $\sk_1 \Ra \{0,1\}^\kappa$ and $\sk_1' \Ra \{0,1\}^\kappa$ are private key component(s) of \enc~and \mac~schemes, respectively (can be a unified key for \grapheneae) . 
				
				\State \Return $\SK_1$
			\end{algorithmic}
			\algrule
			
			\begin{algorithmic}[1]
				\Statex \underline{$\SK_{i+1} \as \grapheneupd(\SK_i)$}: 
				
				\State $\sk_{i+1} \as H(\sk_{i})$, $sk_{i+1}' \as H(\sk_{i+1}')$ \Comment{delete $\SK_i$ from the memory}
				
				\State \Return $ \SK_{i+1} \as (\sk_{i+1},\sk_{i+1}')$ 
			\end{algorithmic}
			\algrule
			
			\begin{algorithmic}[1]
				\Statex \underline{$\sigma_{i,j+1} \as \grapheneagg(\sigma_{i,j},\sigma_{j+1})$} (require $i \le j$)
                    \State \textbf{if} $\bagg=1,2,3$ \textbf{then} aggregate using hash, XOR, or $Add_q$. (see Fig. \ref{fig:graphene})

                \State \Return $\sigma_{i,j+1}$ \Comment{delete $(\sigma_{i,j},\sigma_{j+1})$ from the memory}
			\end{algorithmic}
			\algrule
			\begin{algorithmic}[1]
				
				\Statex \underline{$T_i \as \grapheneoo(\SK_i, n)$:} Precomputation depends on \oo~bits. 
                    \For{$j=i,\ldots, i+n$}
                    \State $r_{j} \as \enc_{s_j}(n)$ where $s_{j} \as \prf_{\sk_i}(j)$ \Comment{delete $s_j$ from the memory}
                    \State $r_j' \as \prf_{\sk_i'}(j)$
                    \EndFor

                    \State $\boldsymbol{\SK_{i+n+1}} \as \grapheneupd(\boldsymbol{\SK_i})$ 
                    \State $T_i \as (T_i^{\texttt{ENC}} \as \{r_j\}_{j=i}^{i+n}, T_i^{\texttt{MAC}} \as \{r_j'\}_{j=i}^{i+n})$
				\State \Return $(\SK_{i+n+1}, T_i)$
				
			\end{algorithmic}
			\algrule
            
			\begin{algorithmic}[1]
				\Statex \underline{$(\{c_{j}\}_{j=i}^{i+n},\sigma_{i,i+n}) \as \grapheneencmac(\SK_i,\{m_{j}\}_{j=i}^{i+n})$}
				
                \For{ \textbf{each} $j= i , \ldots , i+n$}
                    \If{$\bencoo=1$ \textbf{and} $\bmacoo=1$}
                        \State $c_{j} \as m_{j} \oplus r_{j}$ \textbf{and} delete $r_j$ from $T_i^\enc \in T_i$
                        \State $\sigma_j \as \macoomac_{r_j'}(c_j)$ \textbf{and} delete $r_j'$ from $T_i^\mac \in T_i$
                    \Else
                        \State $c_j \as \encstdenc_{\sk_j}(m_j)$ \textbf{and} $\sigma_j \as \macstdmac_{\sk_j'}(c_j)$
                        \State $\SK_{j+1} \as \grapheneupd(\SK_j)$
                    \EndIf
                    
                    \State $\sigma_{i,j} \as \grapheneagg(\sigma_{i,j-1},\sigma_j)$

                \EndFor
				
                \State \Return $(\{c_{j}\}_{j=i}^{i+n},\sigma_{i,i+n})$ 
			\end{algorithmic}
			\algrule
			\begin{algorithmic}[1]
				\Statex \underline{$\{m_{j}\}_{j=i}^{i+n} \as \grapheneverdec(\SK_i,\{c_{j}\}_{j=i}^{i+n},\sigma_{i,i+n})$}

                \State $b \as \macaver(\SK_i, \{c_{j}\}_{j=i}^{n}, \sigma_{i,i+n}) $ 

                \If{$b = 0$} ~\textbf{abort}
                \Else~decrypt $\{c_{j}\}_{j=i}^{i+n}$ using \encstddec~(similar to step 3,6 in $\texttt{EncMac}(.)$)
                \EndIf
				
				\State \Return $\{m_{j}\}_{j=i}^{i+n}$
			\end{algorithmic}
	}}
	
	\caption{Generic \graphene~Framework}
	\label{alg:graphene}
	\vspace*{-4mm}
\end{figure}

\begin{figure*}[ht!]
	\centering
	\includegraphics[scale=0.5]{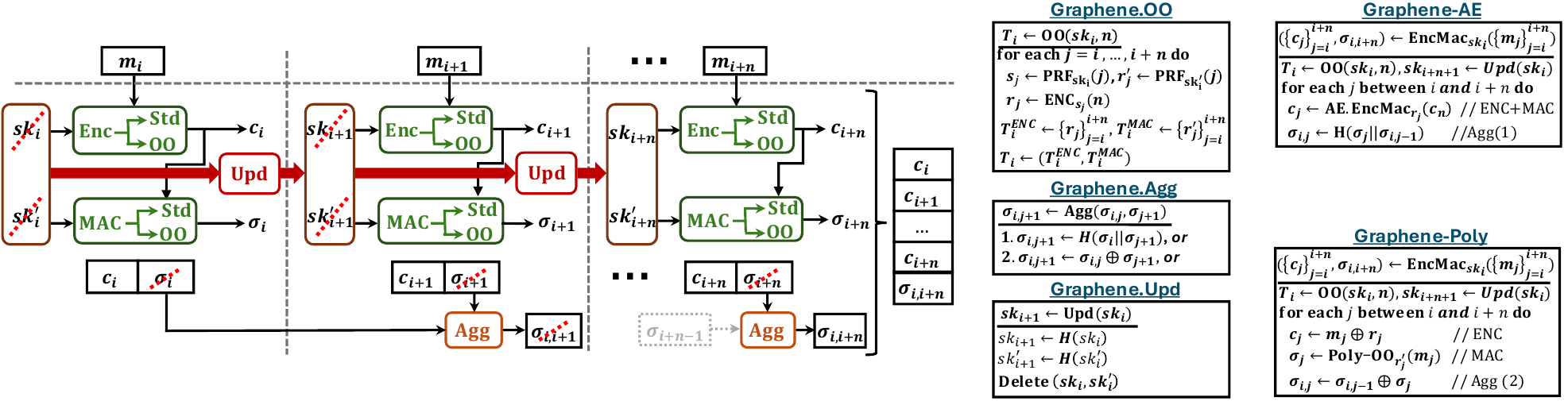}
	\vspace{-2pt}
	\caption{Overview of \graphene~Framework}
	\label{fig:graphene}
	\vspace{-4pt}
\end{figure*}

\subsection{Generic \graphene~Framework} \label{subsec:graphene-generic}

Fig. \ref{alg:graphene} presents the algorithmic description of \graphene.

The key generation (\graphenekg) algorithm takes batch size $n$ as the number of messages to be processed and four flags, first two indicate \enc~and \mac~support for \oo. For example, AES-CTR and Poly1305 \cite{degabriele2024sok} support precomputation with \encoo~and \macoo~properties. \graphenekg~outputs initial private key(s) $\SK_1$ based on the selected primitives. 

We investigate different aggregation algorithms for \grapheneagg:  
{\em(1) Hash-based Accumulator}: computes a digest using a cryptographic hash function. It is an immutable aggregation (i.e., $\sigma \as H(\sigma_1 || \sigma_2)$).
{\em(2) Bitwise XOR}: is an efficient mutable aggregation using XOR operations proportional to the tag size ($\sigma \as \sigma_1 \oplus \sigma_2$). 


The authenticated encryption (\grapheneencmac) algorithm executes \grapheneoo~if \oo~flags are enabled before receiving telemetry. The step 2 in \oo~precomputes a cipher stream of the same size as the input. \oo~can be made input-independent by storing $n$ with the maximum possible input size.  During the online phase, \encmac~either retrieves the precomputed encryption stream $r_j$ and performs an XOR per input size (step 3) or uses \encstd~(step 6). Similarly, the authentication tag $\sigma_{j}$ is generated via \macoo~(step 4) according to \mac~flag (\bmacoo) or \macstd~(step 6). After each tag generation, \grapheneencmac~deletes the pre-ciphertexts and one-time keys (step 3-4), and finally outputs the ciphertexts $\{c_{j}\}_{j=i}^{i+n}$ and aggregate tag $\sigma_{i,i+n}$ per batch.

The decryption algorithm (\grapheneverdec) consists of verifying $\sigma_{i,i+n}$ using aggregate verification (\macaver) where batch verification is used if $\bbver$ is enabled. Certain universal MAC~schemes (e.g., LC \cite{etzel1999square}) support batch verification, which allows efficient tag verification. Otherwise, it generates individual tags and performs aggregation similar to  \grapheneencmac~(Step 2-8). 

\subsection{Graphene Instantiations} \label{subsec:graphene-variants}
Fig. \ref{fig:graphene} depicts a high-level overview of \graphene~with its two main instantiations, described as follows:

\noindent \textbf{\grapheneae.} It employs an AE with key update using a single private key.  We use AES-GCM \cite{mcgrew2004security} (RFC 5288) that offers high efficiency due to hardware acceleration instructions on various platforms (e.g., AES-NI \cite{hofemeier2012introduction}). Moreover, GCM has \oo~with an online encryption time of only XOR operations. We use the first mode of aggregation with SHA-256. \grapheneae~provides: (1) High-level security with standard compliance (e.g., AES-GCM-\{128,256\} with SHA-\{256,512\}), (2) Computational efficiency even in the absence of OO with a single private key, (3) immutable aggregation. \vspace{2pt} 

\noindent \textbf{\graphenepoly.} It uses AES-CTR for \oo~encryption and the universal \mac~Poly1305 \cite{degabriele2024sok} (RFC 7539) for integrity. AES-CTR operates in CTR mode, while Poly1305 consists of a variant of universal MAC: $\texttt{Poly-OO}_{k}(m)=F_{r}(r,m) + s$ where $(r,s)\as \prf_{k}(n)$ is pre-computed offline. We use AES-128 as PRF and SHA-256 for intra-batch and inter-batch key updates, respectively. We adopt the second mode of aggregation (XOR) for optimal efficiency. \graphenepoly~provides: (1) faster running time compared to \grapheneae~while still having standard (RFC) compliance (2) optimal online computation for medium-to-standard ($\kappa=103$-bit) security. Note that \graphenepoly~is extendable to other variants of Poly1305 MACs \cite{degabriele2024sok} with adjustable performance-security and up to  $224$-bit security but still without standard compliance. \vspace{2pt}


\subsection{Security Analysis} \label{subsec:security-discussion}

The security of \graphene~relies on \enc~and \mac~schemes while considering forward security and tag aggregation:

\begin{theorem}
    If \A~can $(t',\epsilon', q_G)$-break \graphene, then one can construct a PPT algorithm $\mathcal{B}$ to $(t,\epsilon,q_H)$-break $H$, or $(t,\epsilon,q_P)$-break  F-IND-CPA-secure \enc, or $(t,\epsilon,q_S)$-break FA-EU-CMA-secure \mac, respectively, where $t' = t + \mathcal{O}(q_H+q_P+q_S)$ and $\epsilon' = \epsilon \cdot i_0$ with $i_0$ being the maximum supported \grapheneencmac~operations. 

\end{theorem}

\begin{proof}
The F-IND-CPA security of \enc~(instantiated with AES-CTR-128) is reduced to its underlying \prf~(i.e., AES-128) and the pre-image resistance of the key update function $H$ (i.e., SHA-256). \A~can $(t, \frac{1}{2^{129}}, q_P)$ and $(t, \frac{1}{2^{128}}, q_H)$-break AES-128 and SHA-256, respectively. The FA-EU-CMA of universal \mac s is reduced to the security of universal hash functions (i.e., GHASH and Poly respectively), \prf, and $H$. \A~can $(t, \frac{1}{2^{128}}, q_S)$ and $(t, \frac{14 \cdot \frac{|m|}{16}}{2^{106}}, q_S)$-break GHASH \cite{mcgrew2004security} and Poly1305 \cite{degabriele2024sok}, respectively, where $|m|$ is the input size. 
%



\graphene~provides fine-grained forward security by updating private keys per data item. Upon  system compromise at $i < j' \le i+n$, $\mathcal{A}$ do not have access to $\{\SK_j \}_{j < j'}$ related to prior generated ciphertexts and tags.  The only available keys to $\mathcal{A}$ are $\SK_{i+n+1}$ output of $H$, and the remaining one-time keys $\{r_j, r_j' \}_{j \ge j'}$ output of $\prf_{\sk_i}$ and $\prf_{\sk_i'}$ in $T_i$. Thus, $\mathcal{A}$~decrypts the traffic associated with $\{\sk_j \}_{j < j'}$ 
imply $(t,\frac{1}{2^{128}},q_H)$-break $H$ or $(t,\frac{1}{2^{129}},q_P)$-break $\prf$. 
\end{proof}

\begin{corollary}
    $\mathcal{A}$ recovering individual authentication tags implies breaking the FA-EU-CMA-secure \mac.
\end{corollary}

\begin{proof}
    \graphene~provides holistic integrity by outputting a sequential aggregate tags over a fixed batch size $n$: $\sigma_{i,i+n} \as \grapheneagg(\sigma_i, \ldots, \sigma_{i+n})$ where $\sigma_j \as {\mac_{r_j'}} (m_j) $, without revealing individual tags $\{ \sigma_j \}_j $. Thus, any mix-and-match aggregate forgery \cite{eikemeier2010history} is detected if the queried batch of messages is not equal to the pre-selected window size $n$.
\end{proof}

\section{PERFORMANCE EVALUATION}
\label{sec:perf_analysis}

\begin{table*}[ht!]
	\centering
	
	\resizebox{0.99\textwidth}{!}{
		\begin{tabular}{|l|@{}c@{}|@{}c@{}|@{}c@{}|@{}c@{}|@{}c@{}|c|c|c|c|c|c|c|}
			\hline
			\multirow{2}{*}{\textbf{Scheme}}  & \multirow{2}{*}{\specialcell[]{\textbf{Storage} \\ \textbf{Overhead}}} &
			\multirow{2}{*}{\specialcell[]{\textbf{Transmission} \\ \textbf{Overhead}}} & \multicolumn{3}{c|}{\textbf{Computational Overhead}} 
			& \textbf{Security} & \textbf{FSec} & \textbf{OO} & \textbf{ImAgg} & \textbf{SC} & \textbf{BC}
			\\ \cline{4-6}
			
			& & & \textbf{Offline EncMac} & \textbf{Online EncMac} & \textbf{Online Verification} & (\textbf{$\kappa$}) & \textbf{} & \textbf{} & \textbf{}& \textbf{} & \textbf{} \\
			\hline \hline
			
			\textbf{Standard FAAE} &$2 \kappa$ & $n \cdot |c| + 2 \kappa$ & $ 2 n \cdot PRF + H $ & $ n \cdot (Enc + HMAC + H) $ & $n \cdot (HMAC+H)$ & 128/256 & \cmark & \xmark & \cmark & \cmark  & \cmark
			\\ \hline \hline
			
			\textbf{\grapheneae} & $\kappa + n \cdot |c|$ & $n \cdot |c| + \kappa$ & $ n \cdot (PRF+AES.Enc_{|m|}) + H $ & $ n \cdot (GHASH + XOR_{|c|} + H) $ & $n\cdot (GHASH+H)$ & 128/256 & \cmark & \xmark & \cmark & \cmark & \cmark
			\\ \hline
						
			
			\textbf{\graphenepoly} & $2 \kappa + n \cdot (\kappa+|c|) $ & $n \cdot |c| + \kappa$ & $ n \cdot (PRF + AES.Enc_{|m|}) + H$ & $ n \cdot (Poly + XOR_{|c|} + XOR_{\kappa}) $ & $n \cdot (Poly+XOR_{\kappa})$ & 103/246 & \cmark & \cmark & \xmark
			& \cmark & \cmark
			\\ \hline
			
			
		\end{tabular}
	}
	\begin{tablenotes}[flushleft] 
		\scriptsize{
			\item $\kappa$ denotes security level. $n$ denotes the batch size. $|m|$ and $|c|$ denote the plaintext and ciphertext sizes. $Add_q$ and $Mul_q$ denote modular addition and multiplication over modulus $q$. {Poly} and GHASH denote one call to Poly1305 \cite{degabriele2024sok} and authenticator of AES-GCM \cite{mcgrew2004security}, respectively. $XOR_\ell$ denotes XOR operations on $\ell$-bit input.  
            FSec, OO, ImAgg, SC, BC denote forward security, offline-online MAC capability, immutable aggregation, standard compliance, and backward compatibility, respectively. 
		}
	\end{tablenotes}
	 
	\caption{Analytical storage and computational overhead of \graphene~variants on batch of input messages}
	\label{tab:analytical_perf}
	\vspace{-2mm}
\end{table*}

Our experimental configuration is as follows:

\noindent \textbf{Commodity Hardware:} We used a desktop equipped with an Intel i9-9900K@3.6 GHz processor and 64GB of RAM.

\noindent \textbf{ARM Cortex M4:} We used STM32F439ZI with 32-bit ARM Cortex-M4 CPU operating at 168 MHz and having 2MB flash memory and 256KB SRAM. It offers cryptographic acceleration (AES-\{128,192,256\}), SHA-256, and HMAC. Cortex-M4 is widely used in IoT due to its low cost and energy efficiency. 

\noindent \textbf{Software:} We used OpenSSL\footnote{\url{https://github.com/openssl/openssl}} and GMP\footnote{\url{https://gmplib.org}} to implement cryptographic primitives and arithmetic operations, respectively, on commodity hardware. We opt for wolfSSL\footnote{\url{https://github.com/wolfSSL/wolfssl/tree/master/IDE/STM32Cube}} to implement both on Cortex-M4 due to its high efficiency and small code size.  

\begin{figure*}[ht!]
	\centering
	
	\begin{subfigure}[b]{0.23\textwidth}
		\centering
		\includegraphics[width=\textwidth]{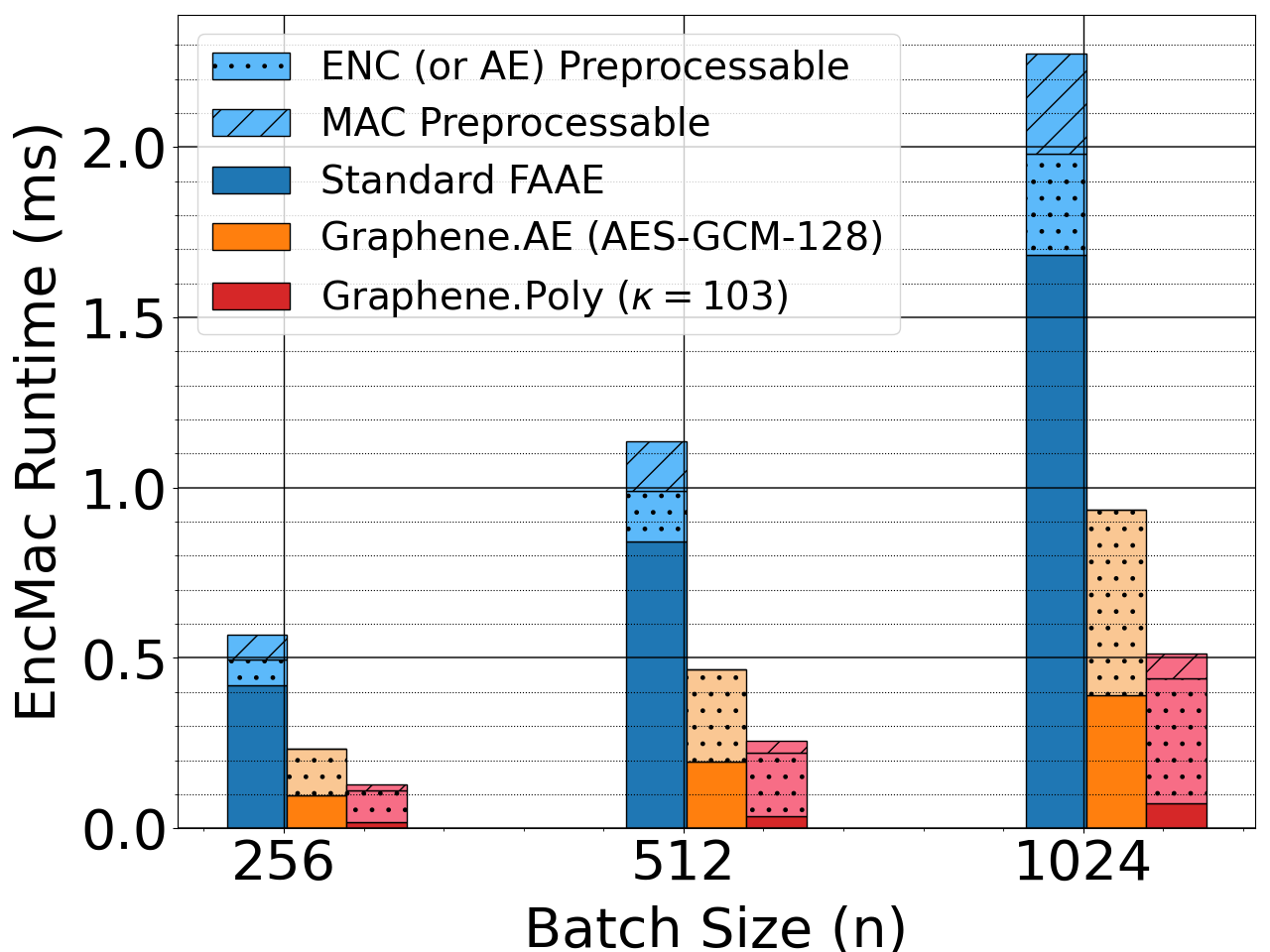}
	\end{subfigure} 
	\hfill
        \begin{subfigure}[b]{0.23\textwidth}
		\centering
		\includegraphics[width=\textwidth]{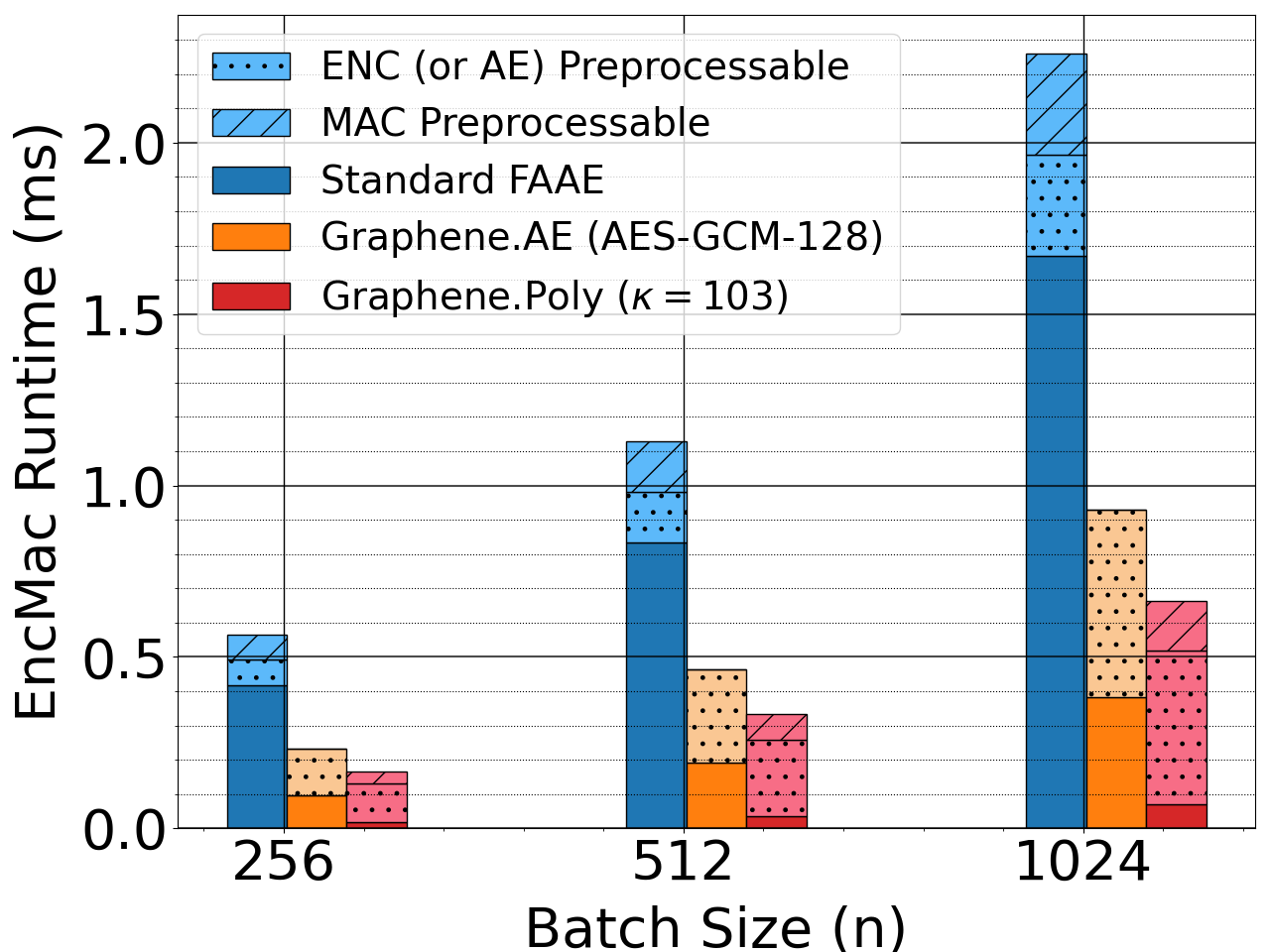}
	\end{subfigure} 
	\hfill
	\begin{subfigure}[b]{0.23\textwidth}
		\centering
		\includegraphics[width=\textwidth]{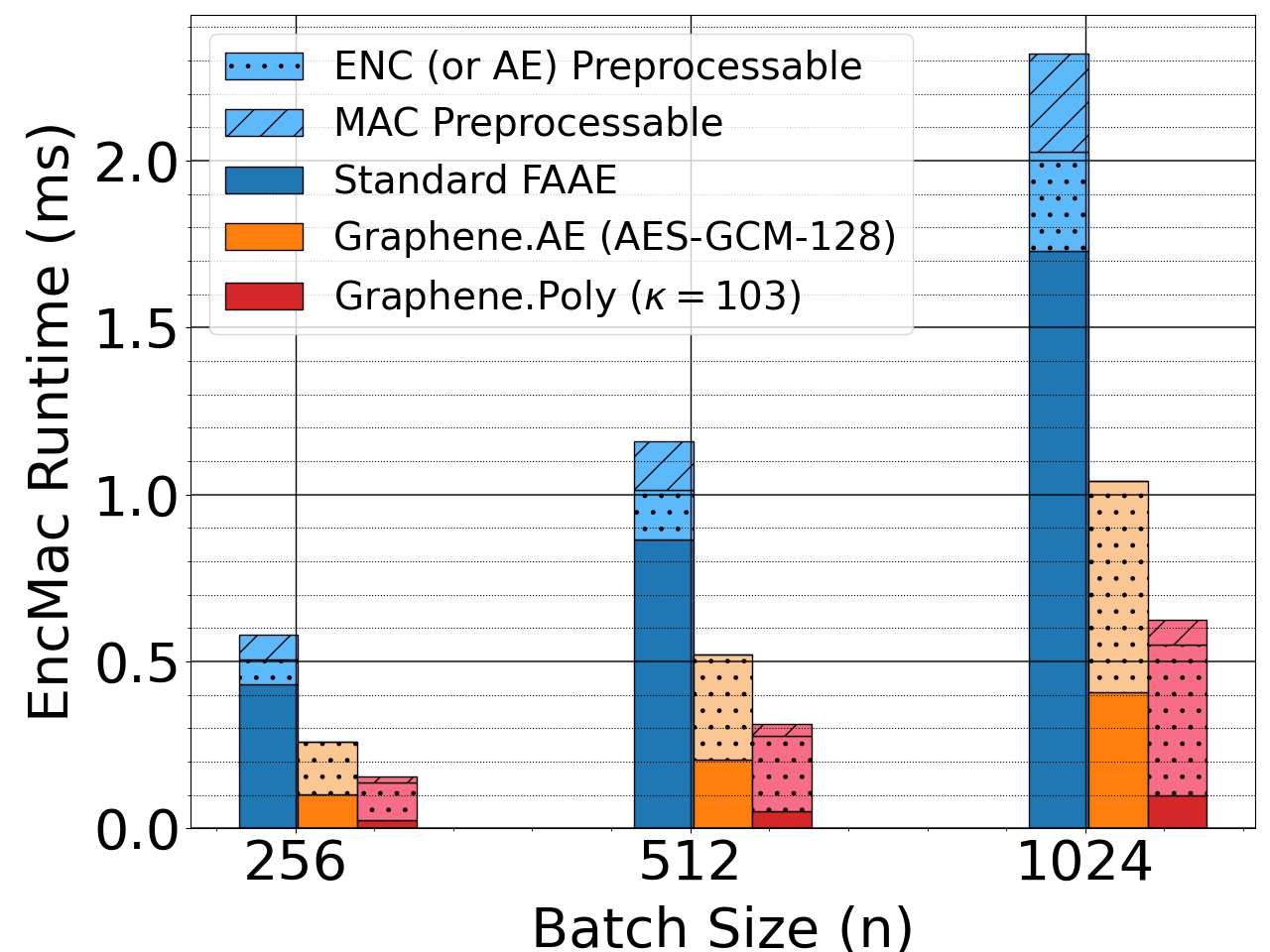}
	\end{subfigure}
        \hfill
	\begin{subfigure}[b]{0.23\textwidth}
		\centering
		\includegraphics[width=\textwidth]{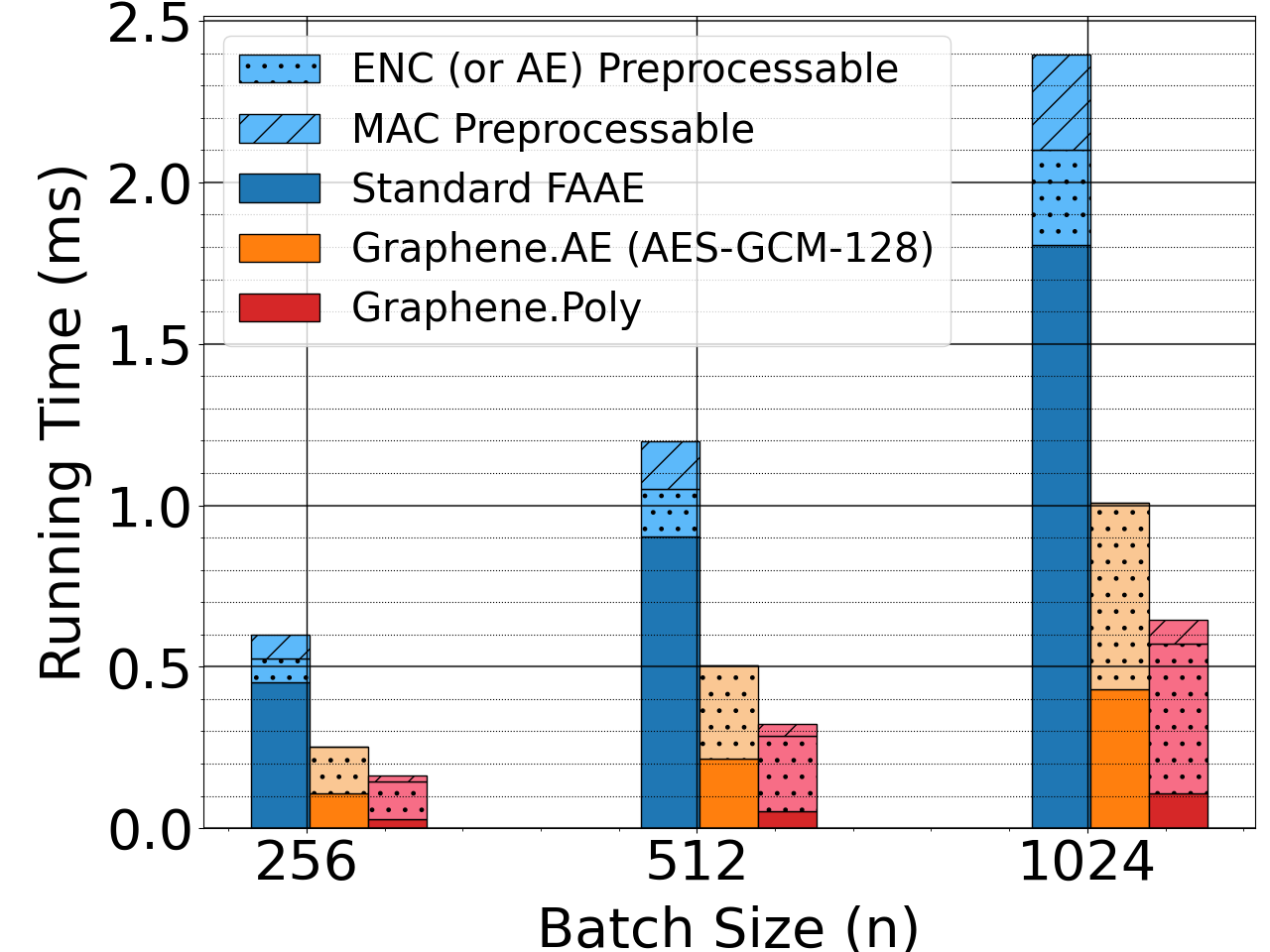}
	\end{subfigure}
	\vspace{2pt}
	
	\begin{subfigure}[b]{0.23\textwidth}
		\centering
		\includegraphics[width=\textwidth]{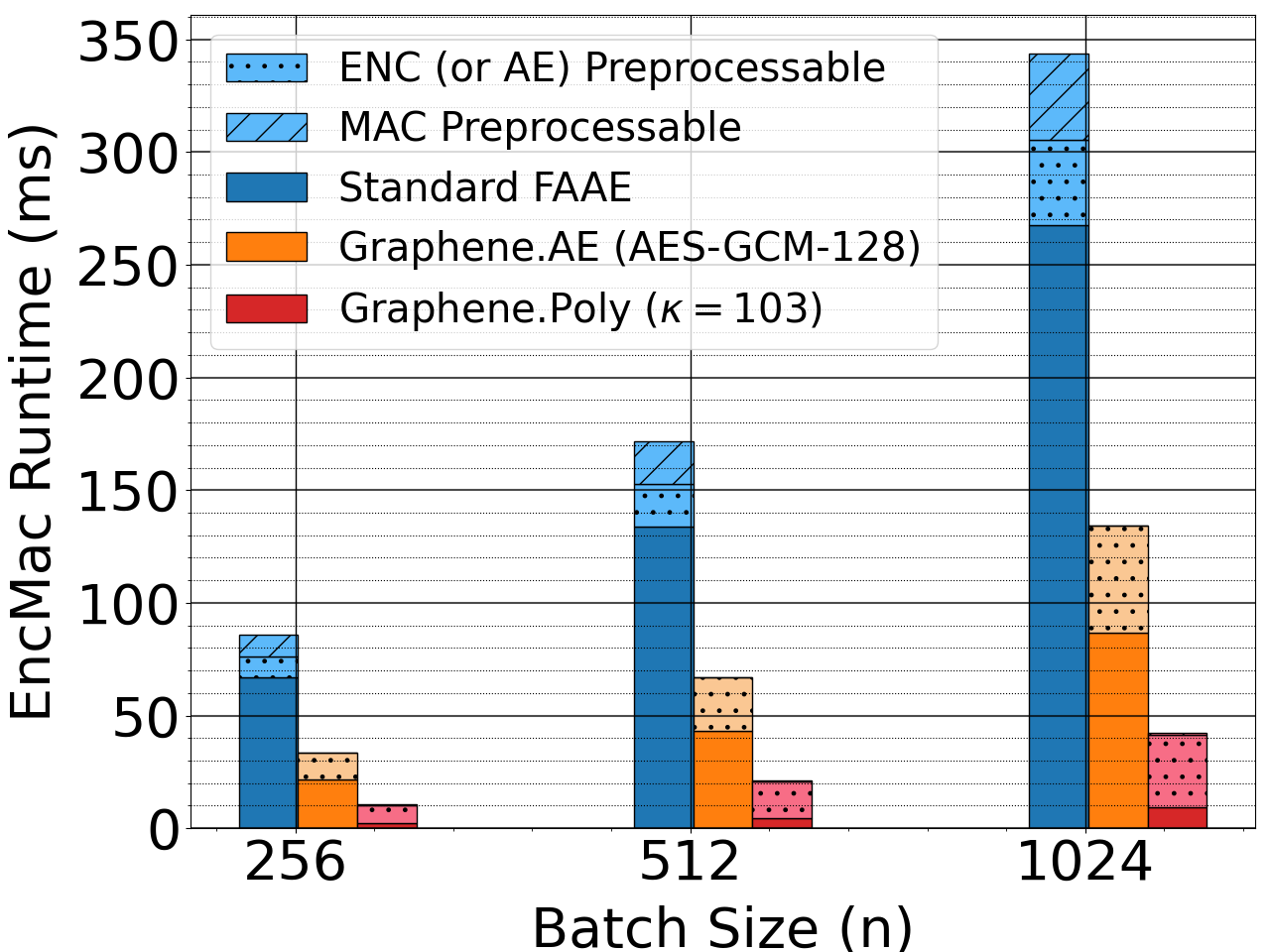}
	\end{subfigure} 
	\hfill
        \begin{subfigure}[b]{0.23\textwidth}
		\centering
		\includegraphics[width=\textwidth]{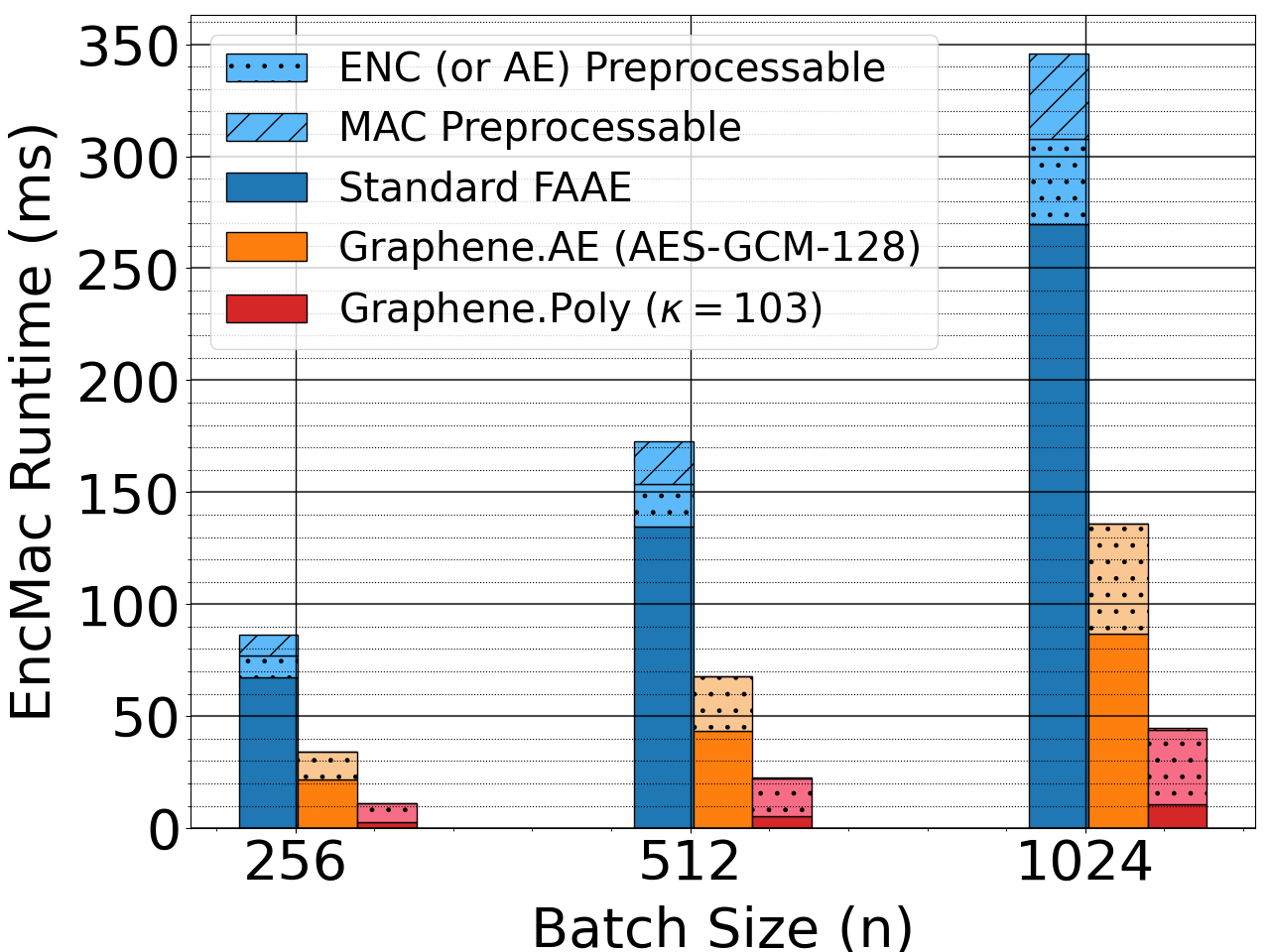}
	\end{subfigure} 
	\hfill
	\begin{subfigure}[b]{0.23\textwidth}
		\centering
		\includegraphics[width=\textwidth]{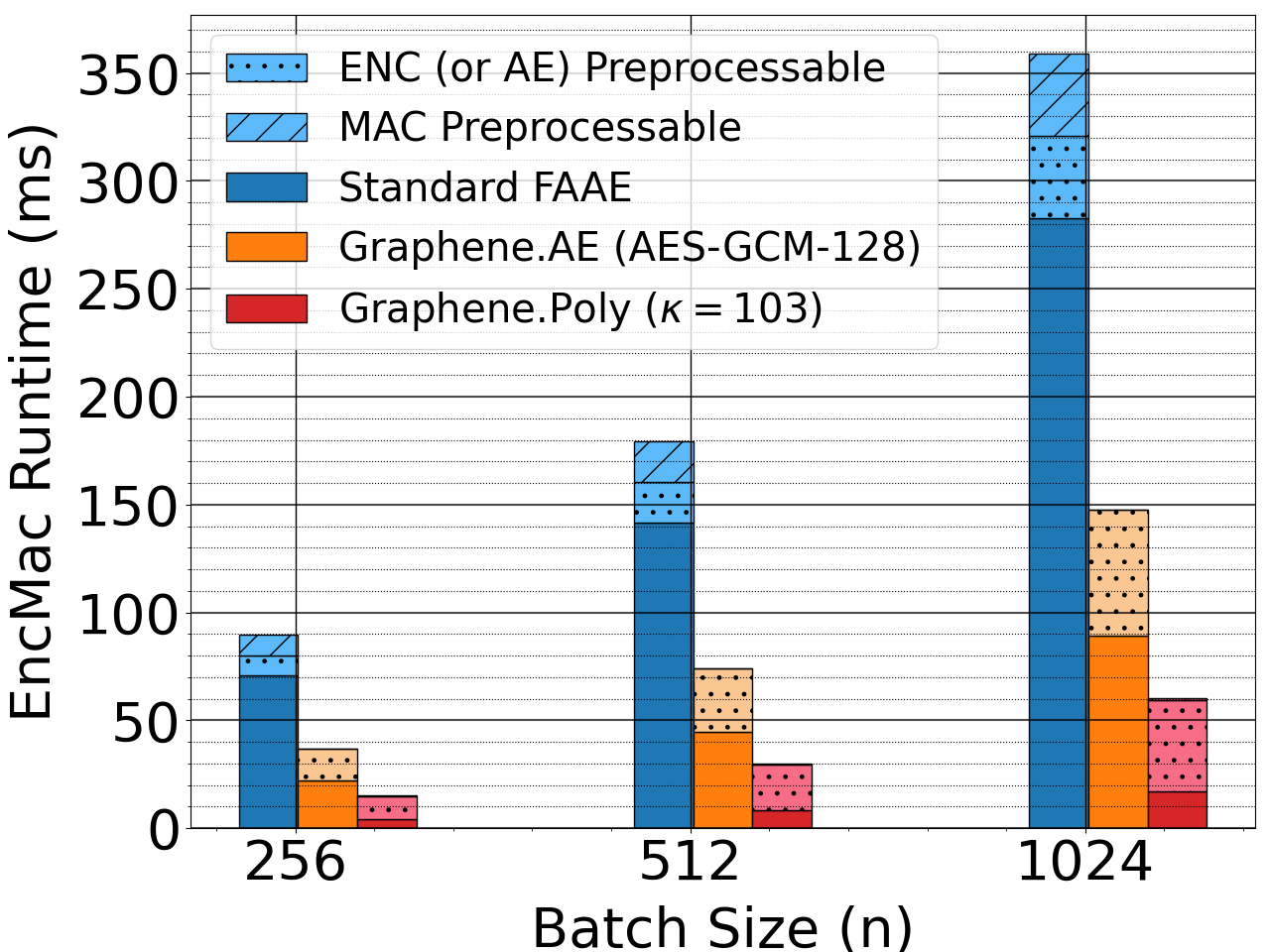}
	\end{subfigure} 
        \hfill
	\begin{subfigure}[b]{0.23\textwidth}
		\centering
		\includegraphics[width=\textwidth]{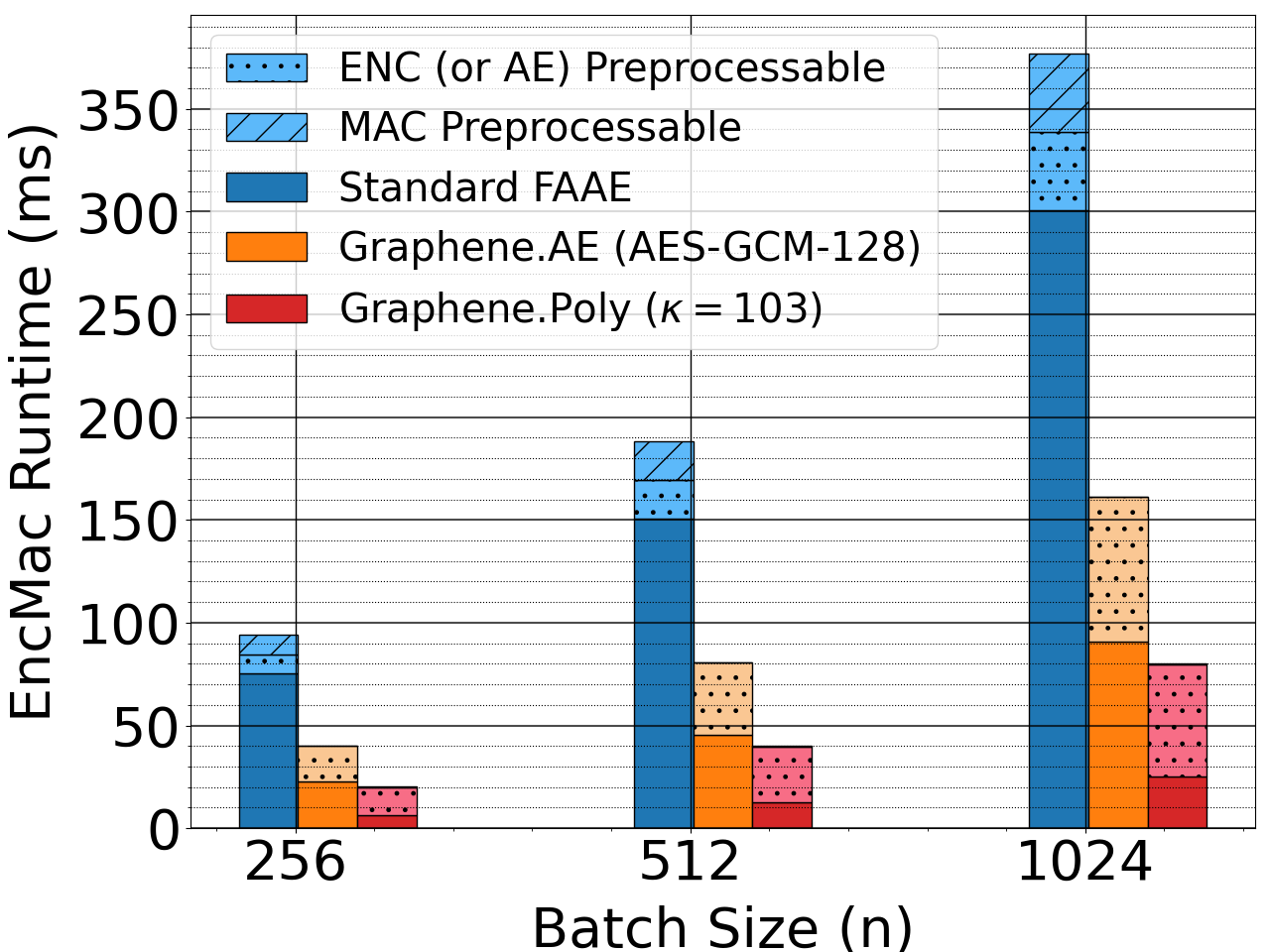}
	\end{subfigure} 
	\vspace{2pt}
	
	\begin{subfigure}[b]{0.23\textwidth}
		\centering
		\includegraphics[width=\textwidth]{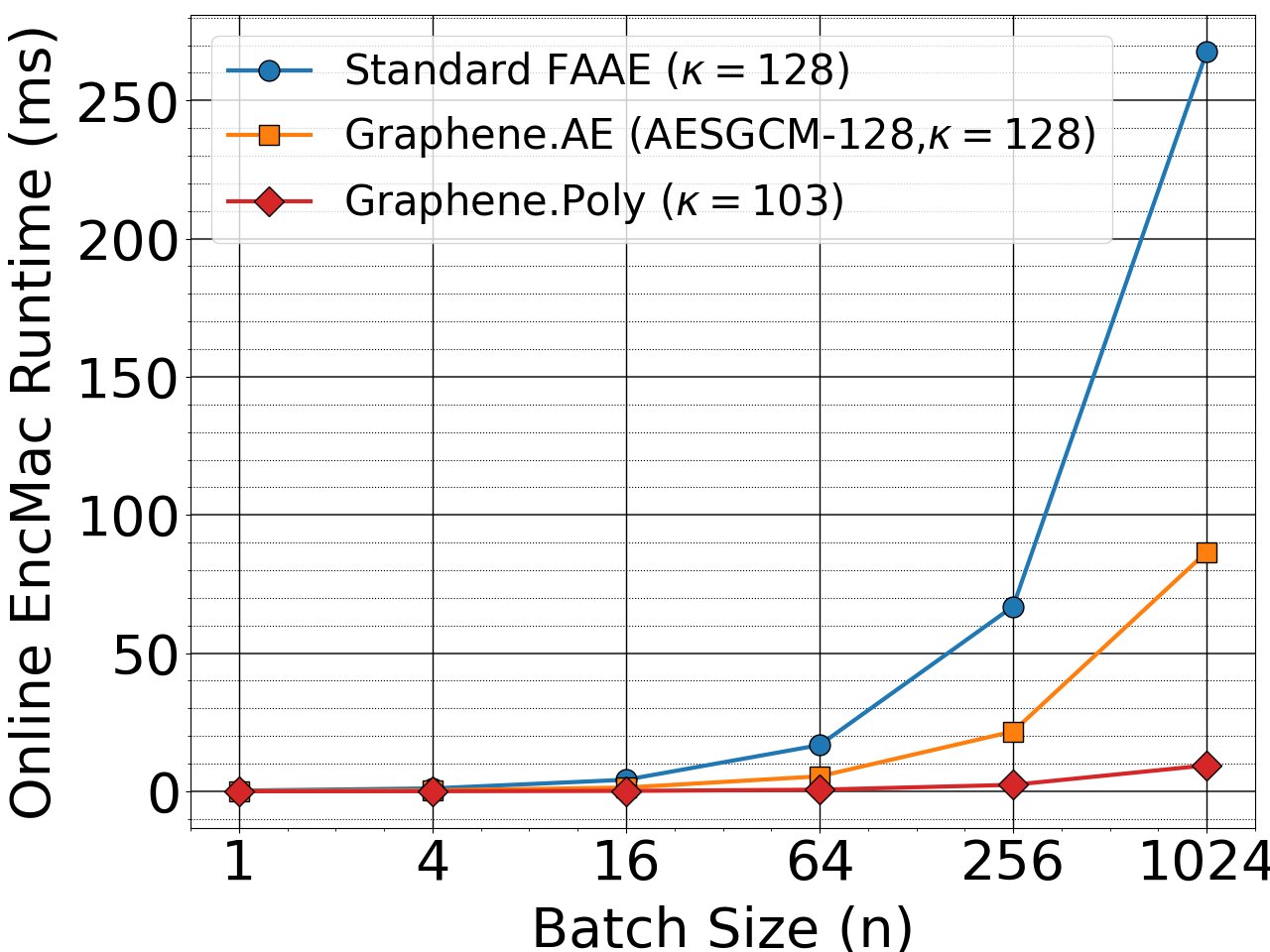}
	\end{subfigure} 
	\hfill
        \begin{subfigure}[b]{0.23\textwidth}
		\centering
		\includegraphics[width=\textwidth]{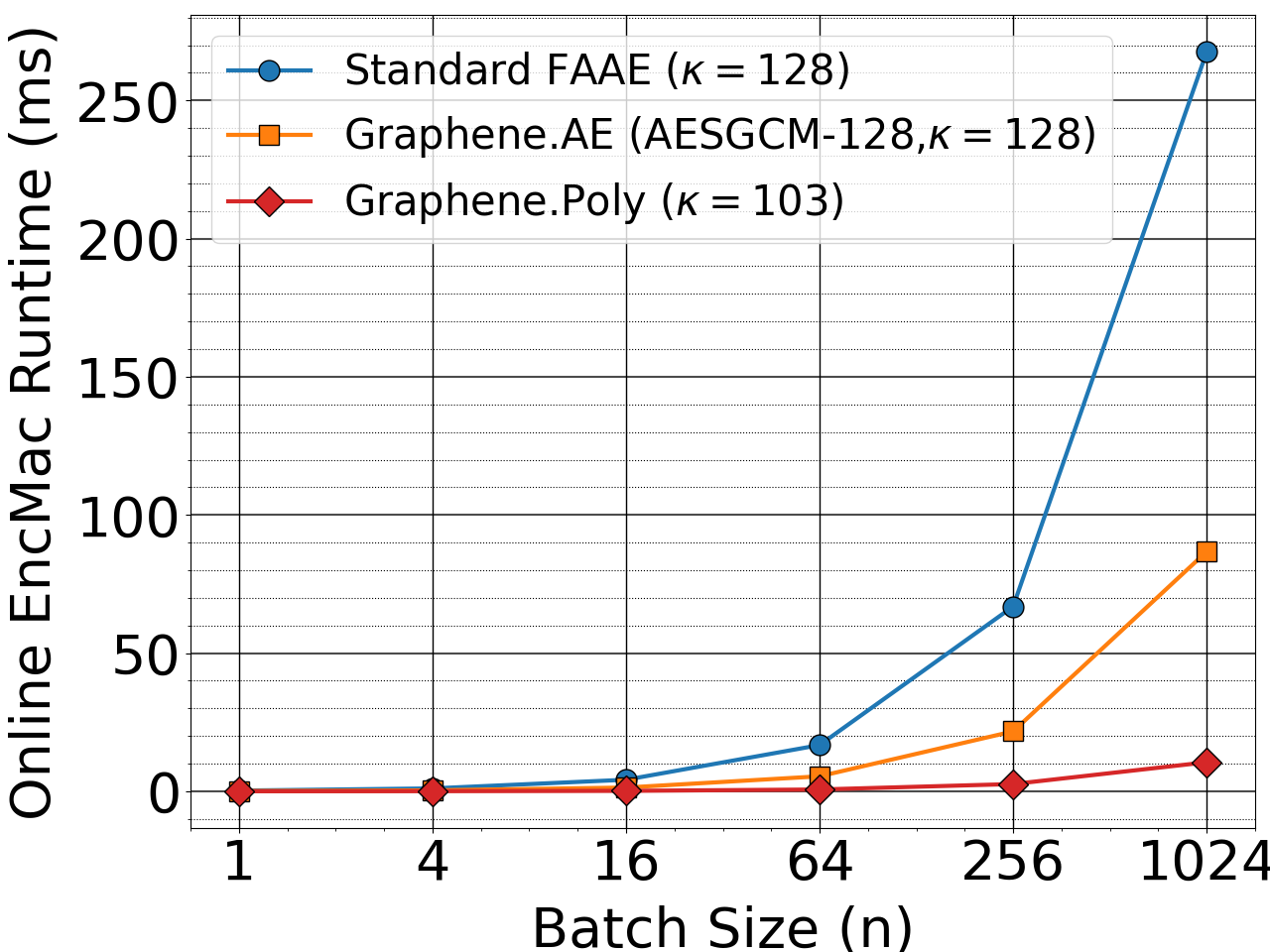}
	\end{subfigure} 
	\hfill
	\begin{subfigure}[b]{0.23\textwidth}
		\centering
		\includegraphics[width=\textwidth]{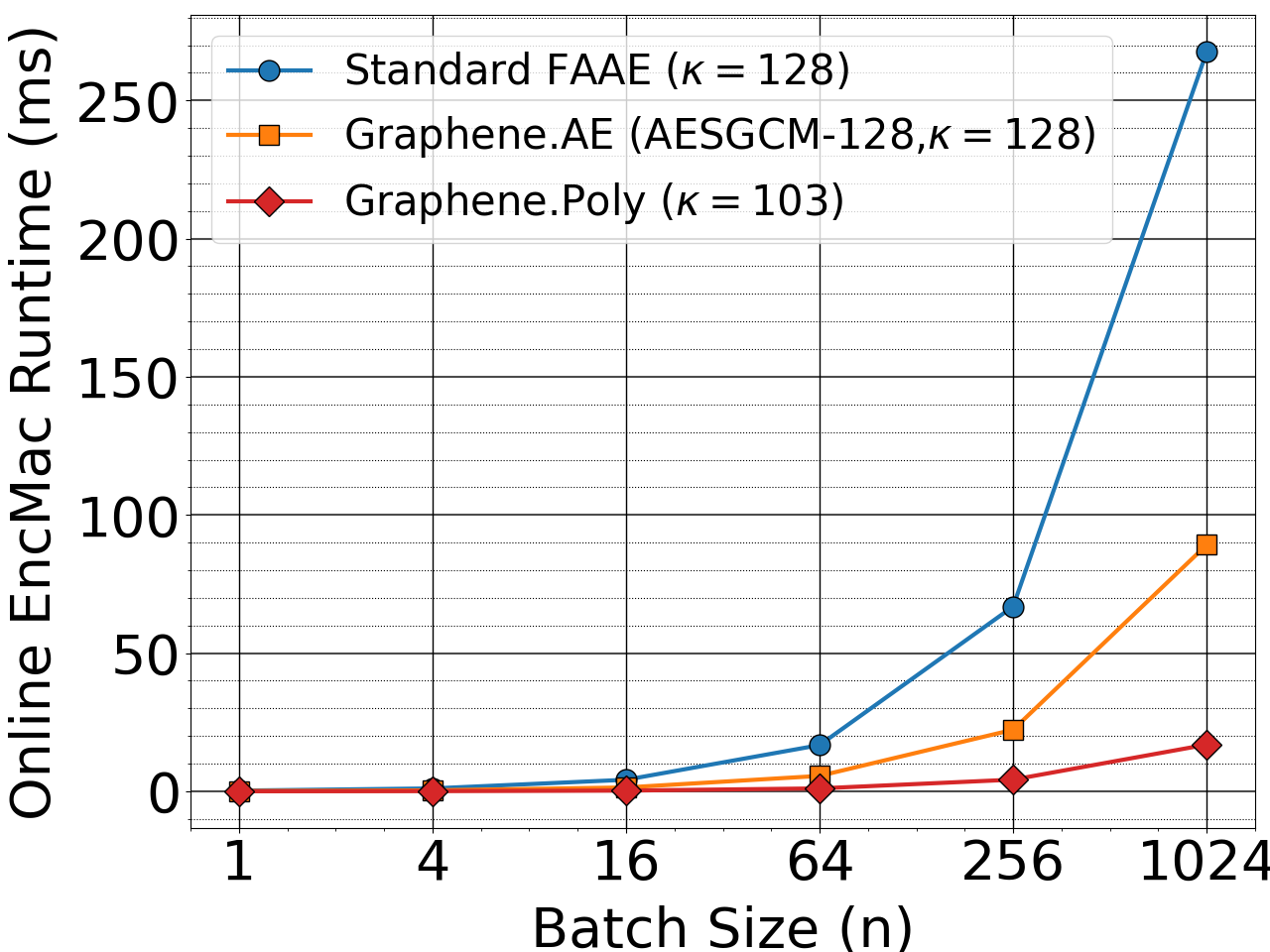}
	\end{subfigure} 
        \hfill
        \begin{subfigure}[b]{0.23\textwidth}
		\centering
		\includegraphics[width=\textwidth]{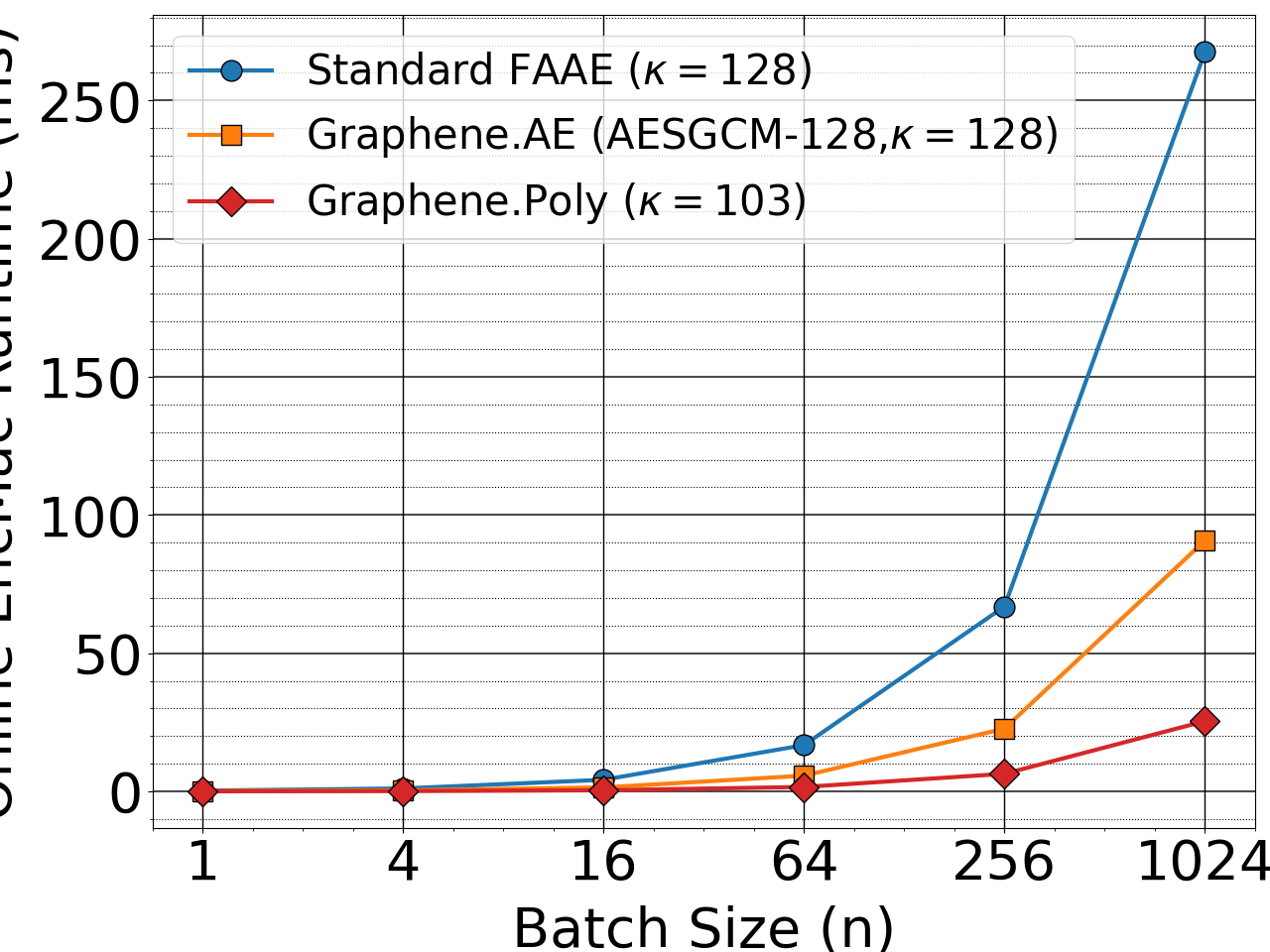}
	\end{subfigure} 
	\hfill
	\vspace{2pt}
	
	\begin{subfigure}[b]{0.23\textwidth}
		\centering
		\includegraphics[width=\textwidth]{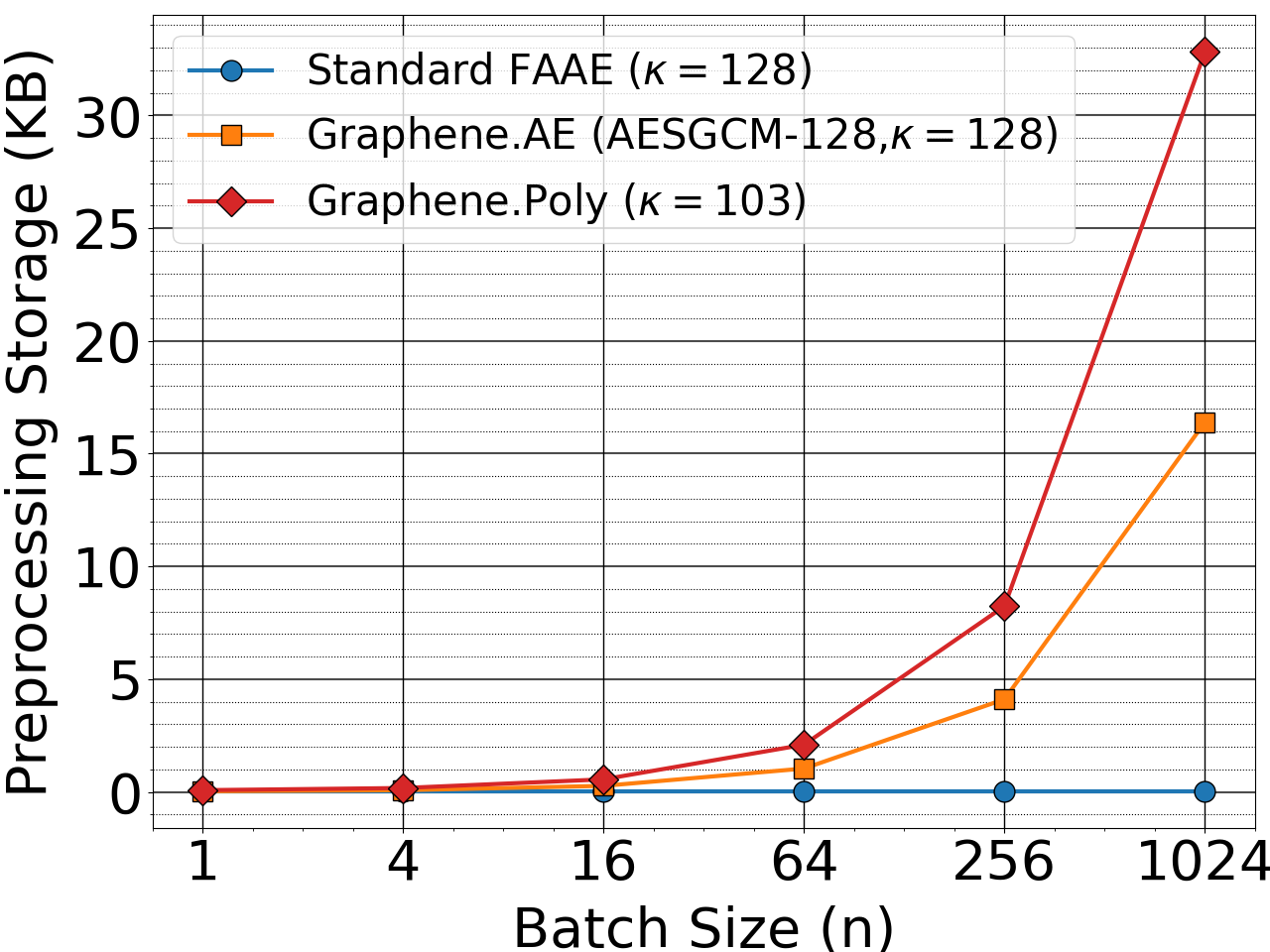}
		\caption{$\boldsymbol{|m|=16 }$ Bytes}
	\end{subfigure} 
	\hfill
        \begin{subfigure}[b]{0.23\textwidth}
		\centering
		\includegraphics[width=\textwidth]{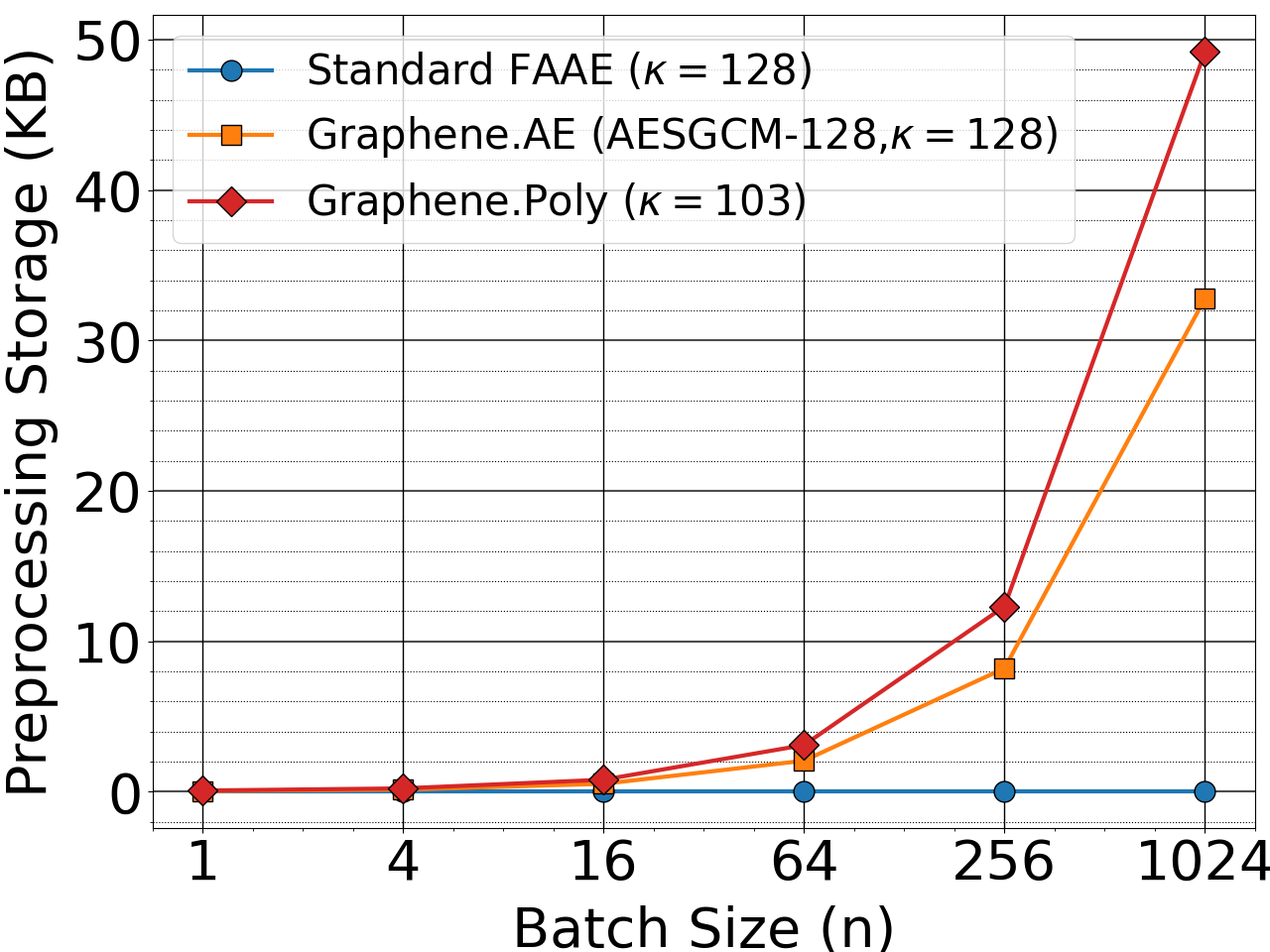}
		\caption{$\boldsymbol{|m|=32}$ Bytes}
	\end{subfigure} 
	\hfill
	\begin{subfigure}[b]{0.23\textwidth}
		\centering
		\includegraphics[width=\textwidth]{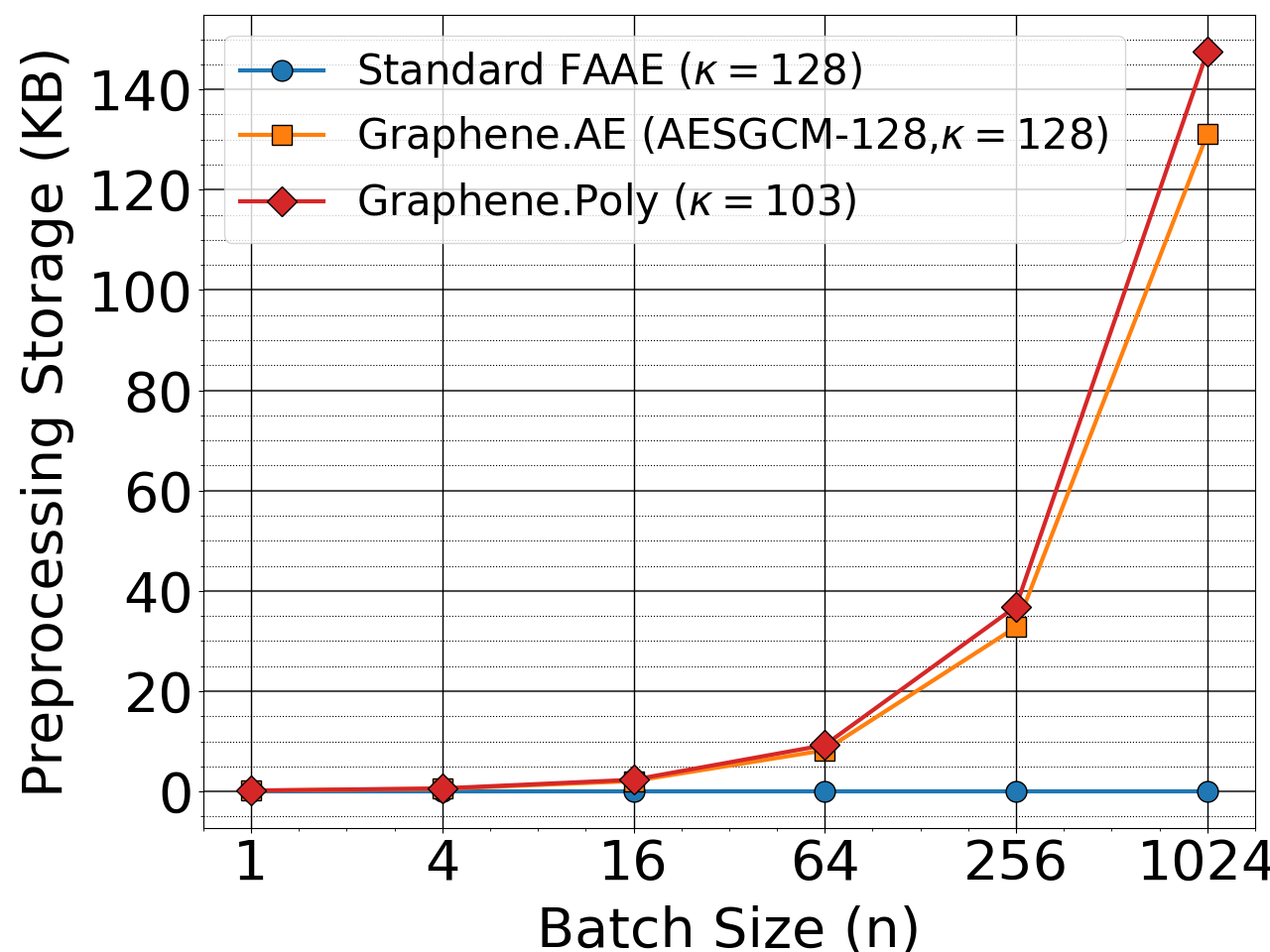}
		\caption{$\boldsymbol{|m|=128}$ Bytes}
	\end{subfigure} 
        \hfill
	\begin{subfigure}[b]{0.23\textwidth}
		\centering
		\includegraphics[width=\textwidth]{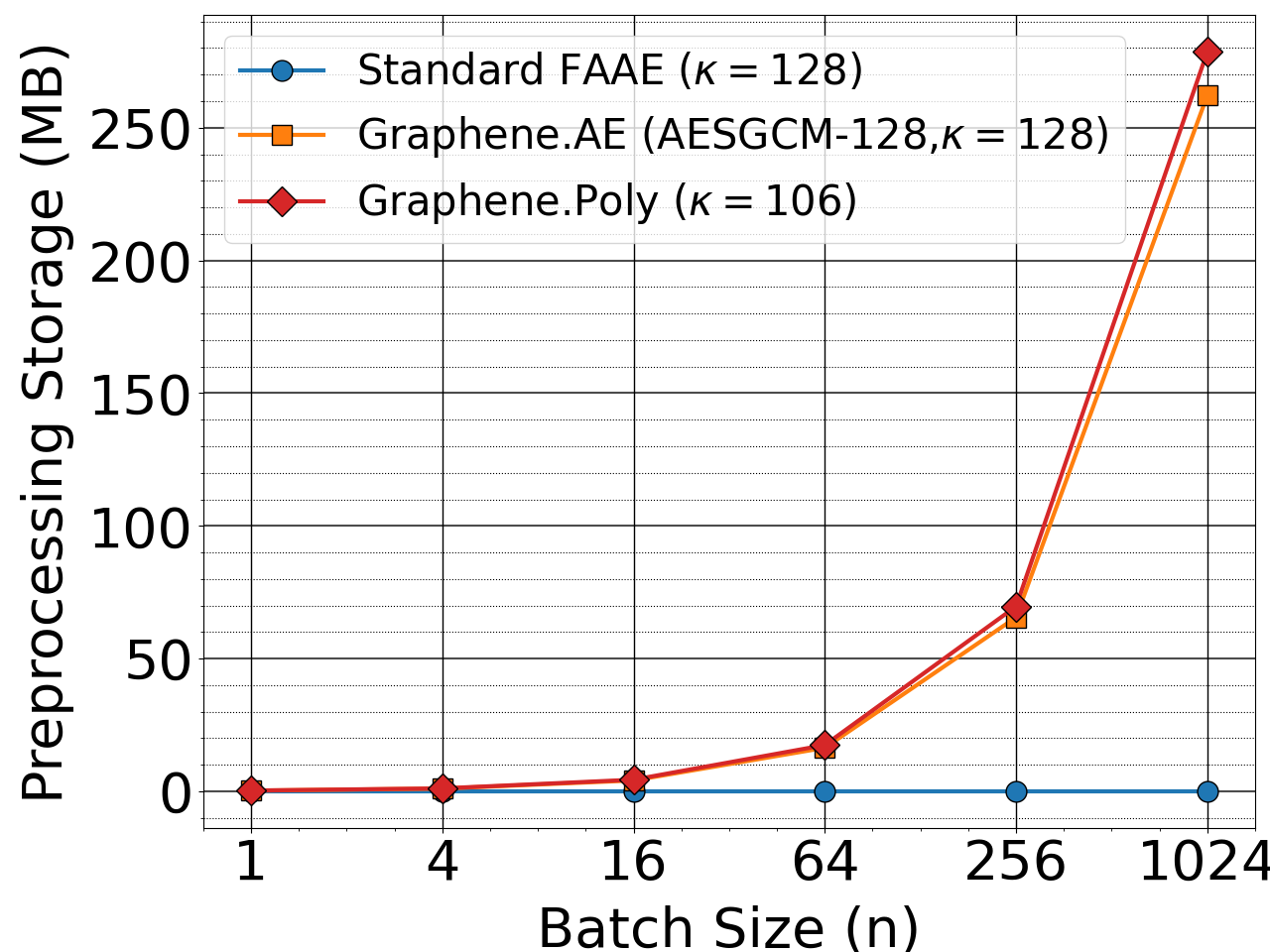}
		\caption{$\boldsymbol{|m|=256}$ Bytes}
	\end{subfigure}

	\caption{OO running time and preprocessing storage overhead of \graphene~variants on commodity hardware (x86/64: $\boldsymbol{1}^{\text{st}}$ row) and 32-bit ARM Cortex-M4 ($\boldsymbol{2}^{\text{nd}}$ and $\boldsymbol{3}^{\text{rd}}$ rows)
	}
	\label{fig:perf-comparisons}
	\vspace{-6pt}
\end{figure*}

Our evaluation metrics are: online computational overhead, tag size, breach resiliency, precomputation storage overhead, and extended properties (e.g., standard compliance).

\subsection{Performance Analysis of Our~Instantiations}
We analyze the performance of \graphene, analytically and experimentally, as shown in Table \ref{tab:analytical_perf} and Fig. \ref{fig:perf-comparisons}. 

We selected a standard FAAE scheme (e.g,~\cite{SUHaSAFSS11}) as our baseline counterpart without precomputation. We used AES-CBC-128~and HMAC-SHA-256 as \encstd~and \macstd, respectively. 
Both of aggregation and key update use SHA-256 as $H$. 
We consider the hardware acceleration of AES, SHA2, and HMAC on both platforms using AES-NI and Cortex optimizations. Our comparison shows the efficiency of UMACs with special properties (without hardware acceleration support) and the benefits of precomputation: \vspace{1pt}

\noindent \textbf{\grapheneae.} It uses AES-GCM with \oo, whose online phase consists only of bitwise XOR operations for encryption but GHASH for authentication.  As depicted in Fig. \ref{fig:perf-comparisons}, on Cortex-M4 and for a batch of 1024 small 16-Byte inputs, the amortized running time is $2.55\times$ faster than standard FAAE, while online being $3.5\times$ faster with additional $16$KB storage. On a batch of 1024 large 128-Byte inputs, the running times become $1.93\times$ and $3.5\times$ faster with $130$KB extra storage, respectively.  As for commodity, for a batch of 1024 small 16-Byte inputs, the amortized and online running times are $2.4\times$ and $4.3\times$ faster than that of FAAE. 
Overall, \grapheneae~offers efficient and scalable performance in offline and online phases while benefiting from hardware acceleration on both platforms. 

\noindent \textbf{\graphenepoly.} It uses AES-CTR-128 and Poly1305 as \encoo~and \macoo, respectively.
AES offers efficient encryption while Poly (RFC7539 standard) is ideal as \macoo~with XOR as the choice of aggregation mode.
Fig. \ref{fig:perf-comparisons} shows that \graphenepoly~achieves the best online running time on both platforms. For example, on Cortex-M4, its overall running time is $8.1\times$ faster than standard FAAE for a batch of 1024 16-Byte inputs, requiring additional storage of only 32KB. Its online running time is $9.23\times$ and $1.35\times$ faster than \grapheneae~on a batch of 1024 16-Byte and 1KB inputs, respectively, with only an additional 16KB. However, it incurs double storage compared to \grapheneae~(i.e., not a standalone AE). Overall, \graphenepoly~is the optimal choice for medium-standard security ($\kappa=106$-bit) with efficient online and overall running time.  \vspace{1pt}


\noindent \textbf{Discussions.} \grapheneae~is best suited for applications requiring high security with efficient online and total running times. \graphenepoly, while incurring some extra storage, is the best choice for efficient online running time for medium-standard security.  Moreover, \graphene~can be easily extended with different MACs to provide additional properties (e.g., \cite{dubrova2018lightweight, wagner2022bp}). For example, on small inputs (1-4 Bytes), BP-MAC \cite{wagner2022bp} achieves the fastest online running time, outperforming Poly1305 by a factor of two on $2$-Byte inputs. \graphene~can also be combined with lightweight secret-sharing schemes for multipath communications \cite{jha2024enhancing} to recover and validate individual messages if aggregate verification fails, thus supporting fine-grained auditability.

\section{CONCLUSION} \label{sec:conclusion}
We present \graphene, a symmetric-key-based FAAE framework designed for resource-constrained IoT systems, addressing critical limitations of existing MAC and AE schemes. By harnessing the synergy between forward-secure key evolution, offline-online cryptographic processing, and aggregate authentication, \graphene~achieves breach resilience, near-optimal online latency, and compact tag generation. Our experimental evaluation on commodity hardware and 32-bit ARM Cortex-M4 microcontrollers demonstrates significant performance gains over existing FAAE solutions, with flexible instantiations to accommodate diverse security and efficiency requirements. We release a full-fledged open-source implementation to promote further research and real-world adoption.

\section*{Acknowledgments}
	This work was supported by the Army Research Laboratory and was accomplished under Cooperative Agreement Number
W911NF-24-2-0078. The views and conclusions contained in this document are those of the authors and should not be interpreted as
representing the official policies, either expressed or implied, of the Army Research Laboratory or the U.S. Government. The U.S.
Government is authorized to reproduce and distribute reprints for Government purposes notwithstanding any copyright notation herein.

\bibliographystyle{ieeetr} 
\bibliography{ref}

\end{document}